\documentclass{jaa}

\usepackage{graphicx}

\begin{document}\sloppy

\title{The Chemical Composition of the Solar Surface}


\author{Carlos Allende Prieto\textsuperscript{1,2}}
\affilOne{\textsuperscript{1}Instituto de Astrof\'{\i}sica de Canarias, V\'{\i}a L\'actea S/N, 38205 La Laguna, Tenerife, Spain.\\}
\affilTwo{\textsuperscript{2}Universidad de La Laguna, Departamento de Astrof\'{\i}sica, 38206 La Laguna, Tenerife, Spain.}


\twocolumn[{

\maketitle

\corres{callende@iac.es}

\msinfo{March 2020}{...}


\begin{abstract}
The Sun provides a standard reference against which we compare the chemical abundances found anywhere else in the Universe. Nevertheless, there is not a unique 'solar' composition, since the chemical abundances found in the solar interior, the photosphere, the upper atmosphere, or the solar wind, are not exactly the same. The composition of the solar photosphere, usually preferred as a reference, changes with time due to diffusion, convection, and probably accretion. In addition, we do not know the solar photospheric abundances, inferred from the analysis of the solar spectrum using model atmospheres,  with high accuracy, and uncertainties for many elements exceed 25\%. This paper gives an overview of the methods and pitfalls of spectroscopic analysis, and discusses the chemistry of the Sun in the context of the solar system.

\end{abstract}


}]


\doinum{12.3456/s78910-011-012-3}
\artcitid{\#\#\#\#}
\volnum{000}
\year{0000}
\pgrange{1--}
\setcounter{page}{1}
\lp{1}

\section{Introduction}

The chemical composition of the Sun, our nearest and most influencing star, is not yet known with high accuracy. The temperatures in the solar interior, as well as in the corona, can reach millions of degrees, and even in the photosphere, where they reach the lowest values, are still thousands of degrees, making it hard to retrieve samples to measure in a laboratory. Our knowledge about the solar chemical composition relies on indirect measurements, through the analysis of oscillations, particles, or light, through physical models of the solar structure.

The abundances derived from optical spectroscopy, which mainly probe the solar photosphere, are usually preferred as indicative of the gas in the protosolar nebula from which our favorite star formed. In the stellar core, hydrogen burning progressively increases the fraction of helium. In the higher atmosphere, significant chemical anomalies are found, most notably the so-called FIP effect, referring to the low abundances found for elements with a high first ionization potential. 


If more stable in time than other regions of the Sun, the solar photosphere is not free from chemical changes. Convective mixing in the solar envelope, which covers the outer 30 \% of the solar radius, continuously cycles photospheric material into regions with much higher temperatures, where lithium is destroyed. At the solar age (about 4500 million years), the photospheric abundance of this element appears depleted relative to its value at birth by a factor of about 150. Models of the Sun indicate that diffusion has reduced, over the solar life time, the photospheric abundance of He by 7\% (0.03 dex), and those of heavier elements by 5\% (see, e.g., Gorshkov \& Baturin 2011). 

Observations from nearby open clusters provide an excellent laboratory to examine these effects. The abundances derived from infrared spectroscopy of the Hyades, for example, show patterns that correlate well with expectations from models including diffusion (Souto et al. 2019). 

In addition, it has been speculated that accretion from circunstellar protoplanetary debris, planets, and material from the interstellar medium, may alter the surface composition of a star over time (Dotter \& Chaboyer 2003). The small abundance differences found between the Sun and a sample of solar twins are strongly correlated with the condensation temperature of the element (Melendez et al 2009; Ram\'{\i}rez et al. 2009). This has been tentatively explained due to refractory material locked down in the solar system rocky planets.

\section{Spectroscopic analysis}

The determination of photospheric abundances from spectroscopy relies on modeling the outer layers of a star, from where light escapes, and solving the radiative transfer through those layers. It is precisely the transfer of radiation, due to the variations of the opacity with wavelength, what gives shape to the observed spectra.

High resolution spectroscopy allows us to sample the physical conditions in the solar photosphere, and infer the chemical abundances in those layers. The assumptions that the solar photosphere is in hydrostatic and radiative equilibrium (with an approximate treatment for the energy exchange associated to convection), and in local thermodynamical equilibrium (LTE), have been extraordinarily fruitful and applied for over a century of research.  With the high quality data available, though, the imperfections associated to these approximations become obvious to the eye.

Convection breaks the hydrostatic equilibrium, distorting and shifting the otherwise nearly perfectly symmetric spectral line profiles. The corrections to the abundances determined from atomic transitions associated to these effects tend to be modest, but hydrodynamical models reproduce line profiles much more accurately. These models also allow us to identify much better overlapping transitions with poorly known transition probabilities, reducing systematic errors. Molecular transitions are, in general, more significantly affected by convection, since the temperature inhomogeneities result in non-linear enhancements of the concentration of molecules in the cooler intergranular lanes.

\begin{figure}[!t]
\includegraphics[width=1.\columnwidth]{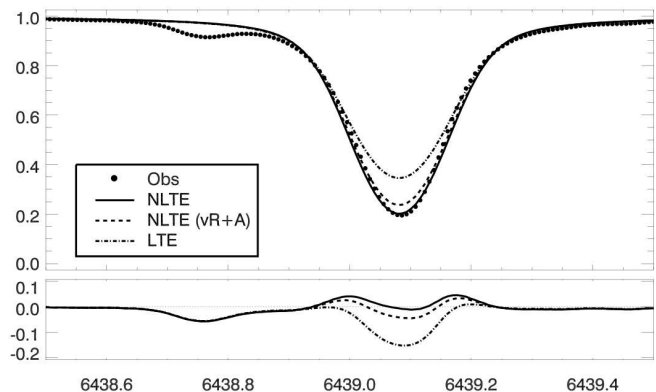}
\caption{Observations (dots) and multiple models for a Ca I transition at 643.91 nm in the solar spectrum. The LTE calculation does not match the line core depth. The two non-LTE calculations differ slightly depending on the source of the electron collisional data. Adopted from Osorio et al. 2019.}\label{f1}
\end{figure}

The sharp variations in opacity produced by lines and continuum edges help to break the local coupling between radiation and matter, and alter the level populations predicted under the LTE assumption. Thus, LTE leads to systematic errors in the abundances inferred from lines of some particular ions, and the usual flux excess predicted for the cores of strong lines, as illustrated in Fig. \ref{f1}.  Departures from LTE are probably the most limiting factor to improve accuracy and precision in the determination of the photospheric abundances of the Sun, and any other star. Without accounting for these departures, it is, in many instances, pointless to upgrade model atmospheres from 1D to 3D.

Recent studies have revealed that departures from LTE in one element can affect those in other elements substantially (Osorio et al. 2020). This complicates modeling, but thanks to rapid progress in the availability of collisional data, performing high-accuracy calculations is now possible. 

Unfortunately, the use of three-dimensional hydrodynamical models in combination with 3D non-LTE radiative transfer is not spreading fast enough. The codes for constructing  hydrodynamical models and the public distribution of  those models are not catching up with the current open-source revolution in science. The same applies to 3D radiative transfer codes. Model atoms for computing departures from LTE are rarely made public. This situation needs to change in the light of the vast data sets of stellar spectra that are currently being collected to study the formation and evolution of the Milky Way and nearby galaxies (APOGEE, GALAH, SDSS, LAMOST, DESI, WEAVE, 4MOST, ...).

Little effort has been devoted to speed up the practical implementation of 3D and departures from LTE in synthetic spectra. If the modeling of 3D effects is (CPU-)time costly, the perturbations they induce on more simple calculations based on hydrostatic LTE models can be characterized and stored for interpolation, speeding up their practical implementation for actual data analysis. This idea was explored in the PhD thesis of Sara Bertran de Lis (2017). On the other hand, there has been significant effort devoted to facilitate and speed up the analysis of observations. For example, the important upgrades in the code SME (Valenti \& Piskunov 1996, Piskunov \& Valenti 2017), The Cannon (Ness et al. 2015), ASPCAP (J\"onsson et al. 2020), or TLUSTY (Hubeny \& Lanz 2017a,b,c).

\begin{figure*}[!h]
\includegraphics[width=1.5\columnwidth]{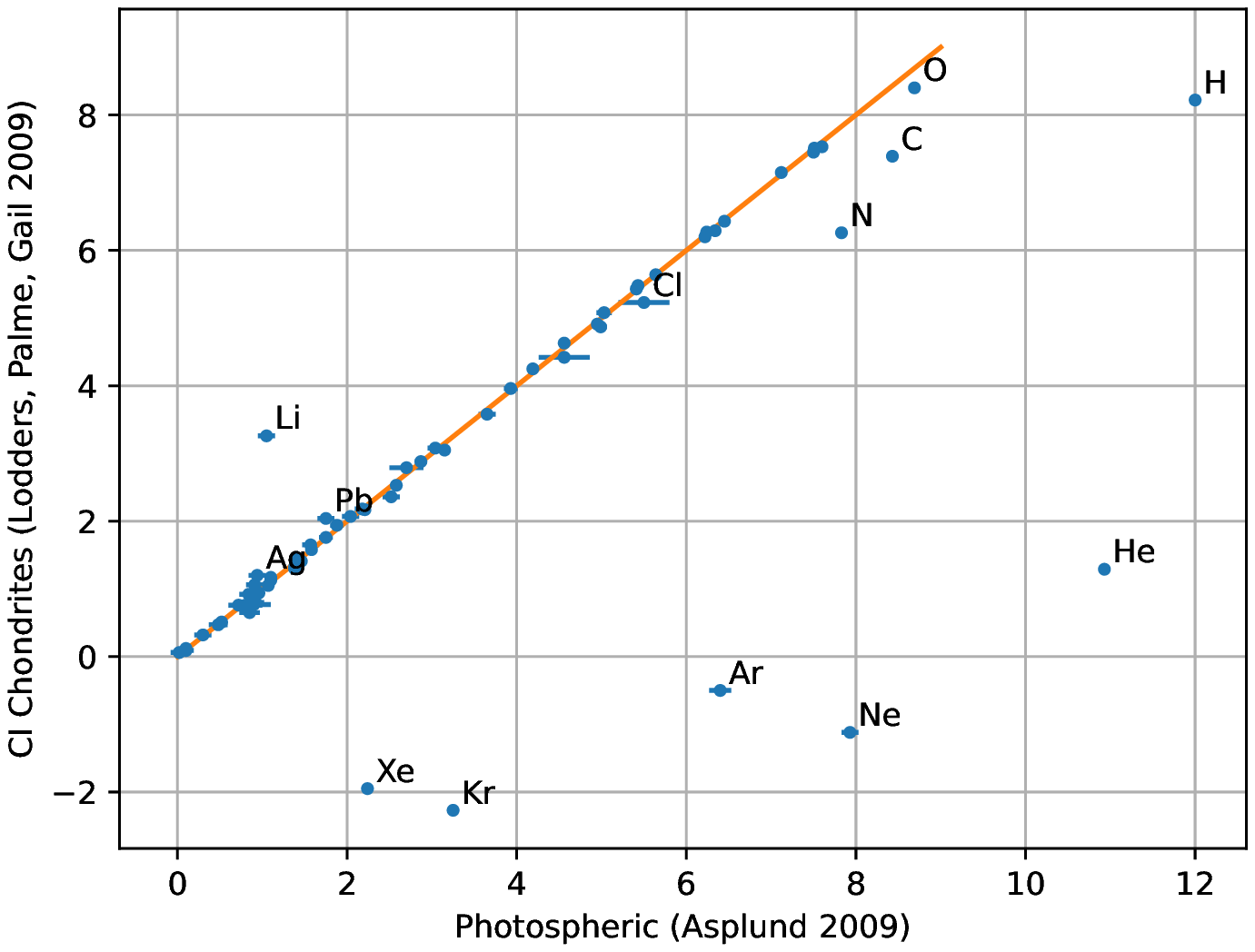}
\includegraphics[width=1.5\columnwidth]{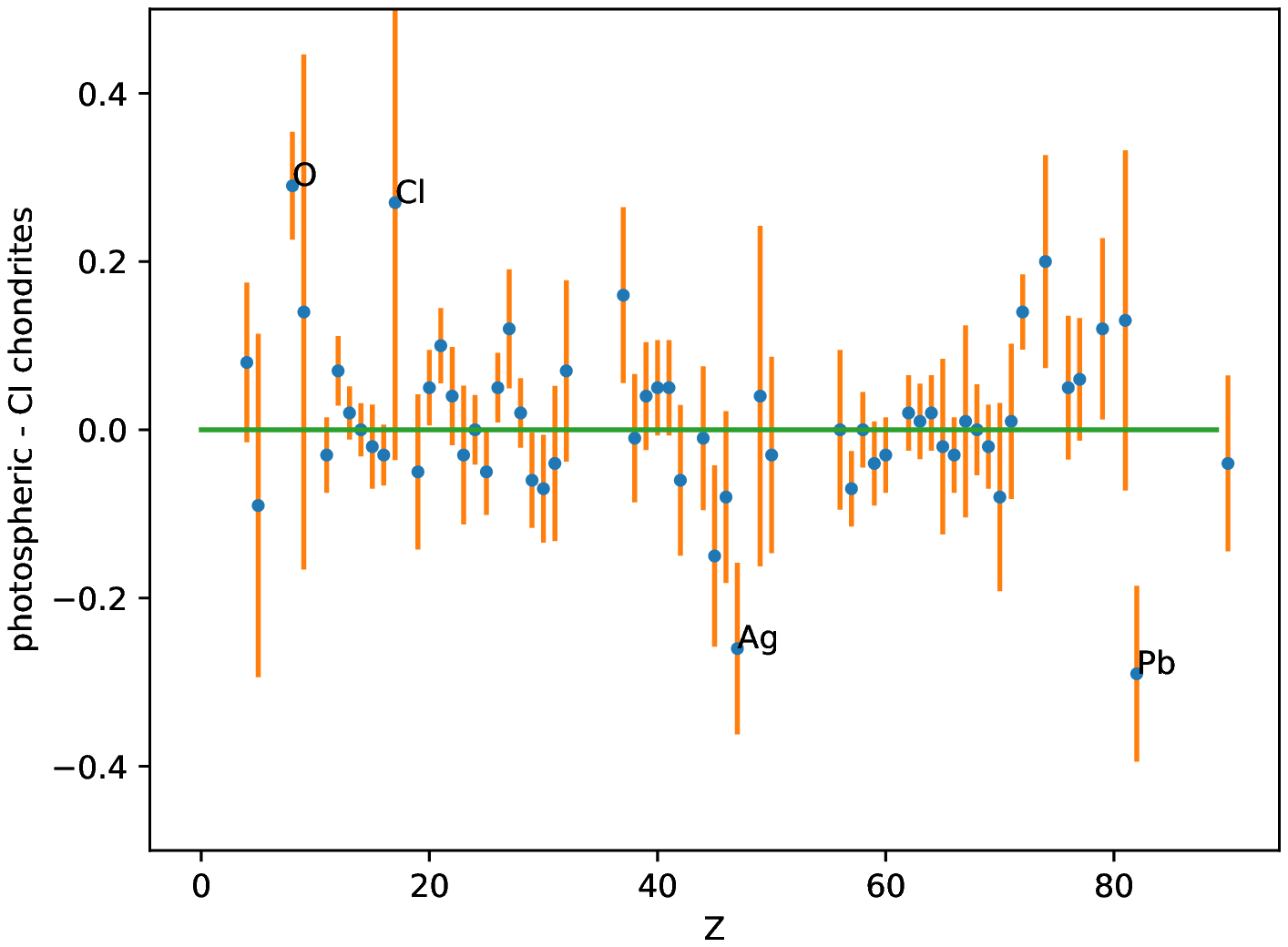}
\caption{{\it Top}: Abundances in Ivuna-type carbonaceous chondrites and in the solar photosphere. {\it Bottom}: Differences between the abundances in the left-hand panel as a function of atomic number, zooming-in to highlight the modest differences found for most elements.}\label{f2}
\end{figure*}

From an observational point of view, there are methodologies that should be discarded for future studies of photospheric abundances. Let's start from the comparison between model and observed spectra. Classical analyses use equivalent widths rather than a direct matching of line profiles. These two strategies differ significantly and may lead to different answers. Equivalent widths give us abundance estimates that are blind, and prone to hide error. For example,  underpredicted collisional damping wings can be compensated by an unknown overlapping transition, preserving the equivalent width, while the comparison of line profiles gives us a measurement of the goodness-of-fit, and therefore  a much higher sensitivity to systematic errors.

In general, few good lines are better than many ok lines. Unknown blends will always induce systematic errors in the same direction -- leaning to higher abundances.  There are many solar atlases available, and sometimes they differ, but the solar optical spectrum is very stable in time. While some studies consider multiple atlases in the determination of abundances, a preferred methodology would try to understand the differences among the data, identify the most reliable source, and stick to that one (or those ones).

\section{Comparison with the early solar system}

Ivuna-type (C-I type) carbonaceous chondrites are considered primitive from a chemical standpoint, and their abundance ratios associated with the early solar system. Despite these meteorites have suffered transformations, the elemental proportions have been preserved, except for the fact that they have lost the volatiles, and therefore they do not provide a good proxy for their original abundances.

\begin{figure*}[!t]
\includegraphics[width=1.5\columnwidth]{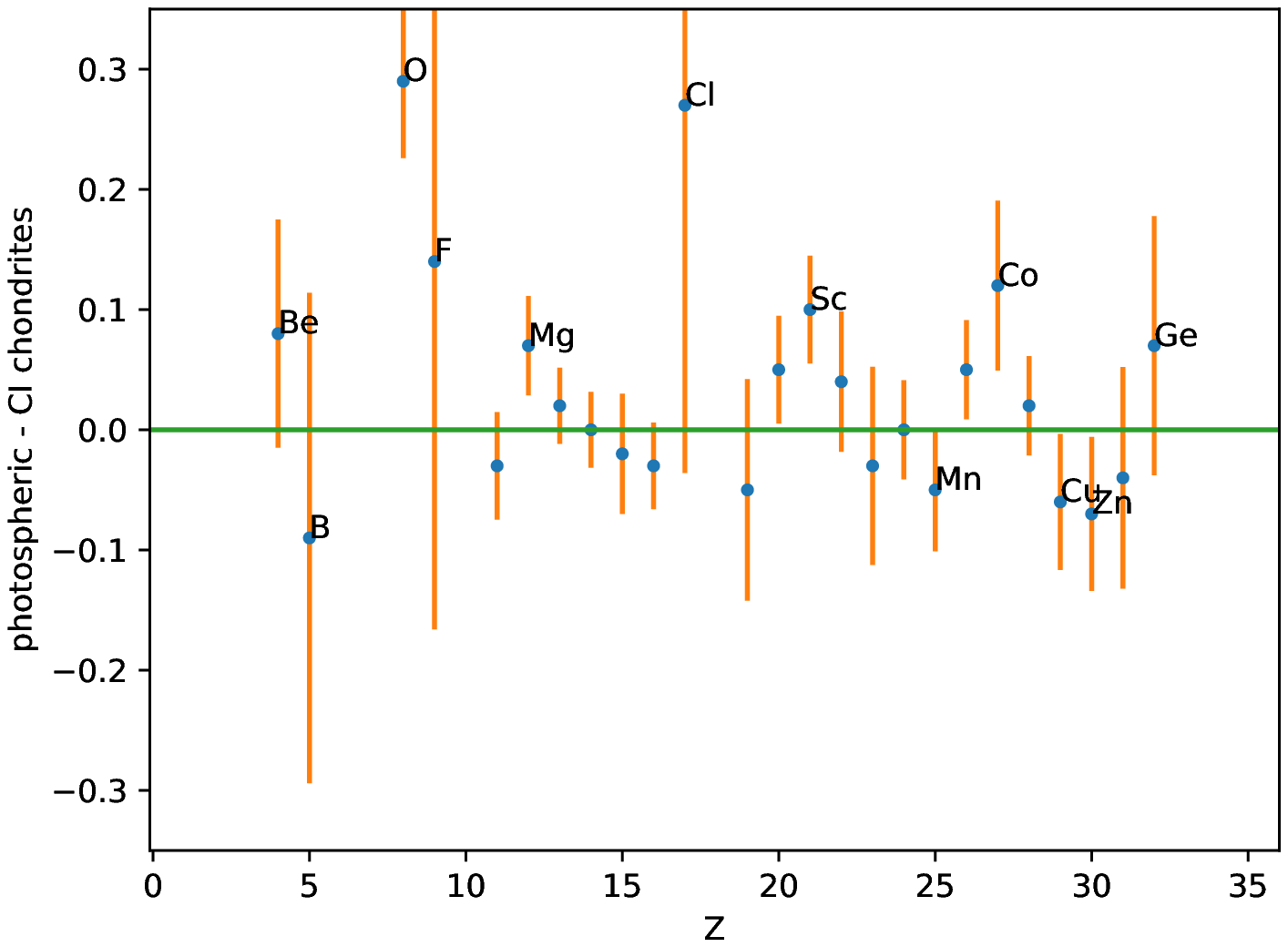}
\includegraphics[width=1.5\columnwidth]{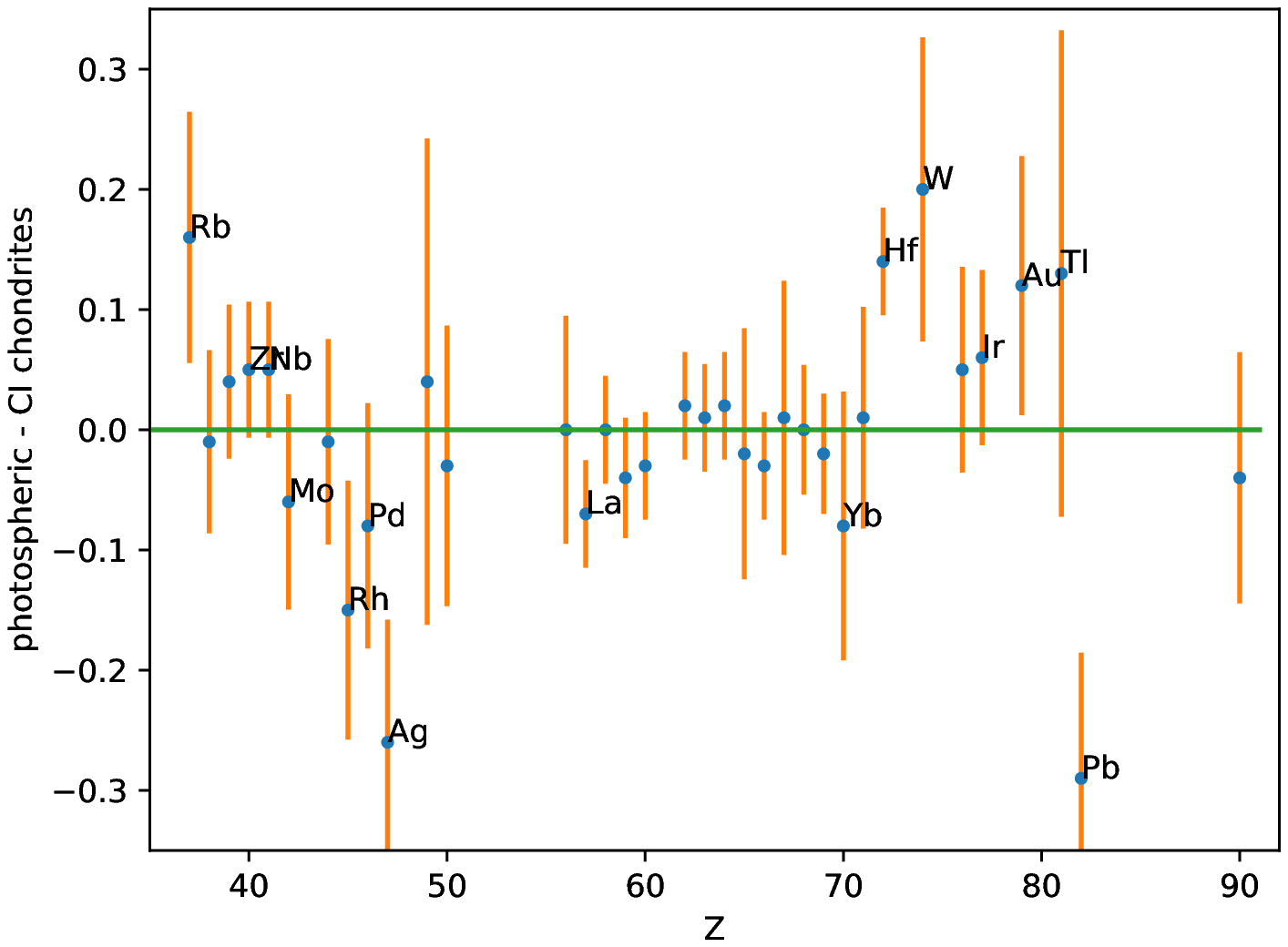}
\caption{Differences between the meteoritic and photospheric abundances. Only elements that shown differences larger than 0.05 dex are labeled.}\label{f3}
\end{figure*}

Good news is that, leaving volatiles aside, there is excellent agreement between the abundance ratios in the solar photosphere and in C-I type chondrites, as illustrated in Fig. \ref{f2} using the data provided by Asplund et al. (2009).  The abundances are  relative to hydrogen and given in the usual logarithmic scale used in astrophysics, forcing hydrogen to have an abundance of 12
\begin{equation}
\log_{10} \epsilon (X) = \log_{10}  \frac{{\rm N}(X)}{{\rm N}(H)} + 12.,
\end{equation}

\noindent where N$(X)$ represents the number density of nuclei of the element $X$. There is no trend in the abundance differences with atomic number, although there is a weak but clear correlation with condensation temperature. Given the absence of volatiles, the abundances are placed on the same scale using silicon (or some times several other refractory elements).

In Fig. \ref{f3} we zoom-in the differences between these two sets of abundances to comment on some interesting features. The beryllium abundances are consistent, but the determination of the solar photospheric value of this element hinges on an ad-hoc continuum opacity added to match OH lines in the vicinity of the Be II resonance lines, at 313 nm (Balachandran and Bell 1998). This aging result should be revised in the light of the most recent continuum opacities, including departures from LTE for magnesium and iron. Boron is another example affected by the uncertainties in the UV opacity, which are likely driving the large error bar apparent in the figure (Cunha and Smith 1999).

The abundances of magnesium are marginally discrepant in our data,  collected from Asplund et al. (2009), but this situation has improved significantly in the more recent literature (Osorio et al. 2020). Similarly, there is a small inconsistency in the calcium abundances, but multi-element non-LTE calculations have been recently shown to solve this issue (Osorio et al. 2020). The same study has shown that the formation of Ca I lines is sensitive to non-LTE effects in magnesium.

Another interesting case is scandium, which shows discrepant abundances between the photospheric and meteoritic values. Zhang et al. (2008) indicate that this elements suffers only small departures from LTE in the solar photosphere. Asplund et al. (2009) underline that there is good agreement in 3D between the abundances inferred for the photosphere from Sc I and Sc II lines, and therefore another close look at this element is warranted. 

The case of cobalt deserves close inspection too. The photospheric and meteoritic values exhibit a 2$\sigma$ discrepancy, which is about the same size and sign of the non-LTE correction adopted for the photospheric value (Bergemann 2008).

Both fluorine and chlorine show large discrepancies, but within the uncertainties of the photospheric values, derived from hydrides observed in sunspots. I would imagine that now is the perfect time to revisit these values, making use of MHD sunspot simulations and three dimensional radiative transfer.

The situation gets more interesting for the heavy elements, where error bars tend to be larger due to complexities in the analysis (e.g hyperfine structure and the need to consider multiple isotopes). These are also affected by the dearth of transitions with reliable data and their tendency to fall in the ultraviolet range, where line blending and an uncertain continuum opacity introduce additional difficulties. Fortunately, there have been significant efforts over the last decade to produce laboratory data for these elements (e.g. Lawler et al. 2007, 2009).

The most discrepant cases are silver and lead, with photospheric abundances lower than the meteoritic ones by 0.25-0.30 dex. These elements are not expected to show such a large discrepancy, which warrants a revision of their photospheric values.

Regarding the uncertainties shown in the differences between meteoritic and photospheric abundances, they cover a wide range, with a typical average of about 20\% (0.08 dex), and mainly reflect the uncertainties in the photospheric values, since the meteoritic ones are more accurate.

Random errors are usually smaller than systematic ones for the photospheric abundances. The latter are the result of poorly measured/computed f-values and damping constants, as well as the approximations discussed in the modeling of the spectrum.

\begin{figure*}[!t]
\includegraphics[width=1.5\columnwidth]{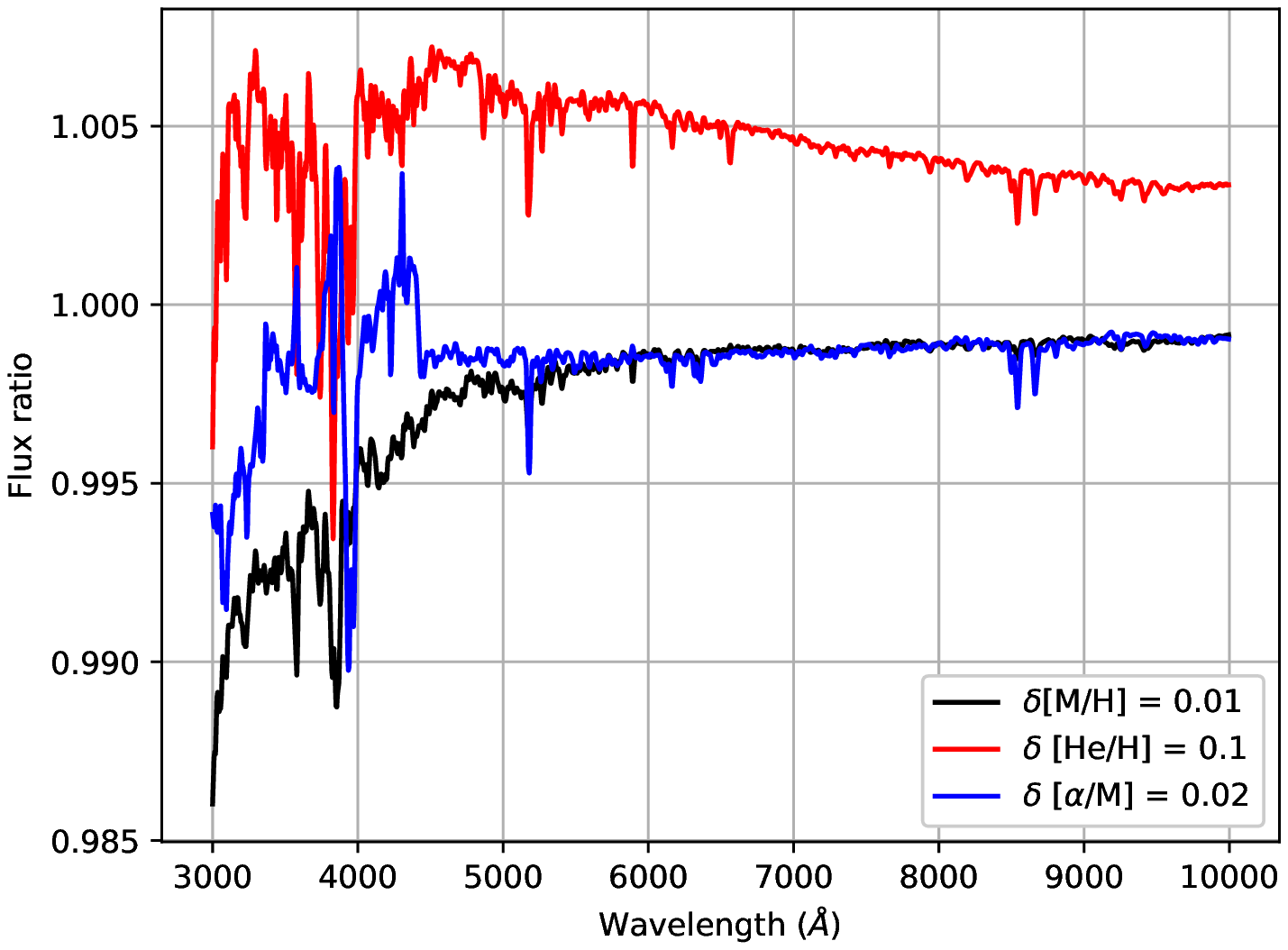}
\caption{Ratio of computed fluxes for solar models with altered compositions relative to nominal (Asplund et al. 2009). The curves correspond to changes of 0.01 dex in overall metal abundance (all metals are increased in sync; black), 0.1 dex in helium abundance (red), and 0.02 dex in the abundances of the $\alpha$ elements (oxygen, neon, magnesium, silicon, calcium and titanium; blue).}\label{f4}
\end{figure*}

\section{Oxygen, carbon and nitrogen}

The solar oxygen abundance is extremely important, given it is the most abundant element after hydrogen and helium. It is as well particularly hard to measure, since there is a limited number of transitions available, atomic lines suffer significant corrections due to non-LTE, and the fact that modeling accurately OH transitions requires taking into account surface inhomogeneities (3D effects).

The reference value in the late 80's was $\log \epsilon$(O) = 8.93 (Anders \& Grevesse 1989), which was subsequently revised to 8.83 (Grevesse \& Sauval 1998), and later proposed to be as low as 8.69 (Allende Prieto et al. 2001) or even 8.66 (Asplund et al. 2005). The most recent analyses accounting for non-LTE and 3D effects stick to similarly low values, around 8.70 (Asplund et al. 2009; Pereira et al. 2009; Scott et al. 2009), although some go a bit higher to reach 8.73-8.76 (Caffau  et al. 2008, 2015).

It has been found that molecular lines agree with atomic lines only when 3D effects are considered. Nevertheless, atomic lines, considering non-LTE effects, give values close to 8.70 both in 1D and 3D atmospheres. Two weak forbidden lines available in the optical have been shown to be immune to departures from LTE and give similar abundances as the permitted atomic lines, although they are blended with overlapping transitions that make the analysis harder.

The situation with carbon is equally complicated. Atomic lines give similar values in 1D or 3D (about 8.44; see Amarsi et al. 2019), but molecular lines (CH, CN, CO) require 3D. As for oxygen, this abundance is about 40\% lower than the value preferred 20 years ago.

The low value in modern analyses of carbon and mainly oxygen has triggered a conflict with models of the solar structure, which prior to the abundance revision enjoyed excellent agreement with helioseismic determinations of the sound speed in the interior of the Sun, and the agreement was dramatically worsened after the abundance reduction (Basu and Antia 2004).

This issue, termed some times as the 'solar crisis' (Ayres 2008), is yet to be resolved convincingly. The opacities used to build stellar structure models are theoretical, and models still use the mixing-length convection for the lack of a fundamental physics-based theory for mixing. Rotating models can explain the discrepancies (Yang 2019), but the community does not seem fully convinced yet that this is the actual root of the problem.

\section{Helium}

Helium contributes about 9\% of all atomic nuclei in the Sun, yet it does not produce observable transitions in the spectrum, except for a chromospheric line at 1050 nm. We are fairly confident about the helium abundance in the Sun from helioseismic data, given their important contribution to the global density, but can the helium abundance be indirectly inferred from the spectrum?

In a recent paper Ting et al. (2018) argue that it is possible to measure oxygen abundances from low-resolution stellar spectra in which oxygen transitions are not actually observed. These authors show the importance of oxygen in the molecular equilibrium in atmospheres of cool stars, which causes the oxygen abundance to affect the strength of molecular bands of molecules that do not contain oxygen. Hema et al. (2020; see also in their contribution in this book) have shown that it is possible to determine helium abundances in red giants through the indirect effect helium has on the strength of Mg I and MgH transitions. 

We have experimented varying the He/H number ratio in the calculation of the solar spectrum from a solar model atmosphere. These calculations use the Sbordone et al. (2007) version of Kurucz's ATLAS9 (Kurucz 2005). The effect is dramatic if feedback to the atmospheric structure is ignored, but when this is considered the impact is much reduced, as illustrated in Fig. \ref{f4}. Despite of the size of the effect, 0.5 \% in flux when He/H is changed by 0.1 dex, this is something worthwhile pursuing, since the spectral signature of this abundance variation is distinct from an overall change in the metal abundances or the alpha elements in particular.

\section{The future ahead}

In the last years synoptic observations of the disk-integrated spectrum of the Sun have been performed with high-resolution stellar spectrographs such as SONG, HARPS, or PEPSI. Relatively recent solar atlases with multiple observations are now available, which constitute a gold mine for testing/improving non-LTE and 3D models.

Plug-and-play radiative transfer and non-LTE codes should become now available thanks to the efforts from several research groups. This will facilitate contributions from different groups willing to work on this important subject.

The makers of hydrodynamical simulations need to publish their models and tools. Similarly to the revolution in astronomy research driven by the public availability of Kurucz's models and codes decades ago, these new tools should trigger dramatic progress, and will do that fast.

Laboratory astrophysics is far from dead, and significant advances have been coming from the groups at NIST, Lund, Imperial College, Wisconsin and other places. We must encourage this work and collaborate with these groups to speed up progress in the field.

If we do our job right the next round of work should give us 0.02-0.04 dex solar photospheric abundances for most elements.

\section*{Acknowledgements}

I thank the organizers for inviting me and congratulate all chemists, terrestrial and cosmic  for the periodic table.

\vspace{-1em}



\end{document}